\documentclass[%
mathleft,%
final,%
]{an}
\usepackage{graphicx}
\usepackage{times}
\begin{document}

% The following seven commands are intended for editorial usage and
% should be ignored by the author(s).
\Pagespan{1}{}% Document's page range.
% If second parameter is left empty, the last page is computed
% automatically.
\Yearpublication{2014}%
\Yearsubmission{2013}%
\Month{1}%
\Volume{335}%
\Issue{1}%
\DOI{This.is/not.aDOI}%

\title{High-resolution absorption spectroscopy of the circumgalactic 
       medium \\ of the Milky Way}

\author{P. Richter\inst{1,2}\fnmsep\thanks{Corresponding author.
  \email{prichter@astro.physik.uni-potsdam.de}}
% Example for footnote, note the usage of the \fnmsep command
% as separator between institute number and footnote mark}
\and  A. J. Fox\inst{3}
\and  N. Ben Bekhti\inst{4}
\and  M. T. Murphy\inst{5}
\and  D. Bomans\inst{6,7}
\and  S. Frank\inst{8}
}
\titlerunning{Absorption spectroscopy of the circumgalactic medium}
\authorrunning{P. Richter et al.}
\institute{
Institut f\"ur Physik und Astronomie, Universit\"at Potsdam,
Haus 28, Karl-Liebknecht-Str.\,24/25, 14476 Golm (Potsdam),
Germany
\and
Leibniz-Institute for Astrophysics Potsdam (AIP), An der Sternwarte 16,
D-14482 Potsdam, Germany
\and
Space Telescope Science Institute, 3700 San Martin Drive, 
Baltimore, MD 21218, USA
\and
Argelander-Institut f\"ur Astronomie, Universit\"at Bonn,
Auf dem H\"ugel 71, 53121 Bonn, Germany
\and
Centre for Astrophysics \& Supercomputing, Swinburne
University of Technology, Hawthorn, Victoria 3122, Australia
\and
Astronomisches Institut, Ruhr-Universit\"at Bochum (RUB), 
Universit\"atsstrasse 150, 44780 Bochum, Germany
\and
RUB Research Department Plasmas with complex interactions
\and
Department of Astronomy, Ohio State University, 140 West 18th Avenue, 
Columbus, OH 43210-1173, USA
}

\received{XXXX}
\accepted{XXXX}
\publonline{XXXX}


\abstract{%
In this article we discuss the importance of high-resolution
absorption spectroscopy for our understanding of the distribution
and physical nature of the gaseous circumgalactic medium (CGM) that 
surrounds the Milky Way. Observational and theoretical
studies indicate a high complexity of the gas kinematics and an extreme 
multi-phase nature of the CGM in low-redshift galaxies.
High-precision absorption-line measurements of the Milky Way's gas 
environment thus are essential to explore fundamental parameters of
circumgalactic gas in the local Universe, such as mass, chemical
composition, and spatial distribution. We shortly review
important characteristics of the Milky Way's CGM and discuss recent
results from our multi-wavelength observations of 
the Magellanic Stream. Finally, we discuss the potential of studying 
the warm-hot phase of the Milky Way's CGM by searching for extremely weak
[Fe\,{\sc x}] $\lambda 6374.5$ and [Fe\,{\sc xiv}] $\lambda 5302.9$ 
absorption in optical QSO spectra.
}

\maketitle

\section{Introduction}

Observational and theoretical studies demonstrate that the
circumgalactic medium (CGM) is a key component in the 
on-going process of galaxy formation and evolution. The CGM
is usually defined as diffuse neutral and ionized gas 
that is located within the virial radius of a galaxy, but 
outside of its (main) stellar body. The gas circulation
in the CGM is governed by the accretion of metal-poor gas from the 
intergalactic medium (IGM), providing fuel for star-formation
in galaxies, and the expulsion of metal-enriched 
material as part of outflows and galactic winds, the latter
process being a {\it result} of star-formation activity.
Consequently, large-scale feedback processes, which are
regulated mostly by the physical conditions in the gas, determine 
the gas distribution in the CGM and its kinematics.

In view of the extreme multi-phase nature of the CGM
(that has typical gas temperatures and densities in the 
range $T=10^2-10^7$ K and $n_{\rm H}=10^{-5}-10$ 
cm$^{-3}$, respectively), absorption spectroscopy
towards extragalactic background sources
represents a particularly powerful method to study 
the nature and distribution of circumgalactic gas 
in the local and distant Universe. This is because
of the presence of a large number of diagnostic lines 
of low, intermediate, and high metal ions 
in the ultraviolet (UV) and at optical wavelengths.
High-resolution absorption spectra of quasi-stellar 
objects (QSOs) and other extragalactic point sources 
therefore provide a wealth of information on the physical 
and chemical properties of the different gas phases
in the CGM of the Milky Way and other galaxies
(e.g., Wakker et al.\,1999; Sembach et al.\,2003; 
Richter et al.\,2001a,2001b,2009,2011; Tripp et al.\,2003;
Fox et al.\,2005,2010; Collins et al.\,2009; Shull et al.\,2009; 
Prochaska et al.\,2011; Tumlinson et al.\,2011a; 
Lehner et al.\,2013; Keeney et al.\,2013; Stocke et al.\,2013).

Most CGM absorption-line studies at low redshift are 
limited to a single QSO sightline that passes the virial
radius of a foreground galaxy (e.g., Thom et al.\,2011; 
Tumlinson et al.\,2011b; Ribaudo et al.\,2011). 
Only for very nearby galaxies,
the angular size of the circumgalactic region is expected to
be large enough to have {\it several} sufficiently bright 
background QSOs per galaxy available that can be used for a 
multi-sightline analysis of circumgalactic gas. Because of
these statistical restrictions, the Milky Way, M31, and other 
galaxies in the Local Group represent important test objects 
to investigate the properties of circumgalactic gas absorbers 
along a large number of randomly distributed sightlines and 
to link the observed sky distribution of the different CGM gas 
phases in the Local Group with the statistical properties of 
QSO absorption-line systems (Richter 2012; Rao et al.\,2013).

\section{Quasar absorption spectroscopy of the Milky Way's 
circumgalactic medium}

\subsection{Kinematics of local circumgalactic gas}

Absorption and emission measurements indicate that
the Milky Way's circumgalactic gas spans a LSR velocity
range between $-500$ and $+500$ km\,s$^{-1}$ (e.g., Wakker
et al.\,2004). The usual classification scheme of the Milky 
Way's circumgalactic gas features includes the so-called 
``high-velocity clouds'' (HVCs) and ``intermediate-velocity 
clouds'' (IVCs). We here define HVCs as gaseous 
structures at $|b|>30$ deg (located outside the Galactic disk)
that are observed in 
H\,{\sc i} 21cm emission or in absorption against an 
extragalactic background source at high radial velocities, 
$|v_{\rm LSR}| > 100$ km\,s$^{-1}$ (see Wakker \& van Woerden 
1997; Richter 2006 for reviews on HVCs). Extra-planar clouds 
with somewhat smaller radial velocities in the range 
$|v_{\rm LSR}|=50-100$ km\,s$^{-1}$ are referred to as IVCs.

% ---------------- Fig 1
\begin{figure}
\includegraphics[angle=0, width=\linewidth]{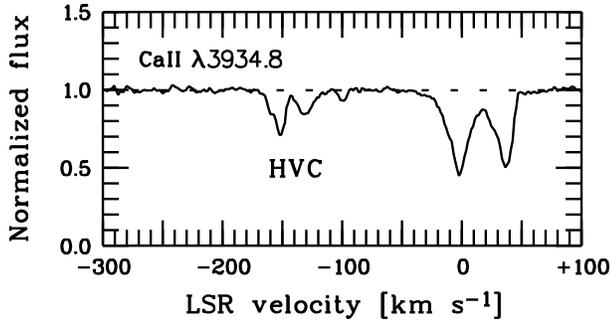}
\caption{Velocity profile of Ca\,{\sc ii} $\lambda 3934.8$ 
absorption in the VLT/UVES spectrum of the QSO PKS\,1448$-$232. 
Absorption from a high-velocity cloud (HVC) is visible near
$v_{\rm LSR}=-150$ km\,s$^{-1}$. 
}
\label{Flabel1}
\end{figure}

Optical absorption spectroscopy of halo stars that span a
distance range of several kiloparsecs in the direction of 
known 21cm HVCs imply that most (if not all) of the HVCs 
are located within $\sim 60$ kpc of the Milky Way 
(e.g., Wakker et al.\,2007,2008; Thom et al.\,2006,2008;
Lehner et al.\,2011,2012). 
HVCs thus represent gaseous structures that are located in 
the inner and outer Galactic halo well within the 
virial radius of the Milky Way. 

Using high-resolution ($R>40,000$) optical spectra of QSOs 
and other extragalactic background sources obtained with 
the Ultraviolet and Visible Echelle Spectrograph (UVES) 
installed on the Very Large Telescope (VLT), we have studied 
the distribution of neutral gas and its kinematics in the 
Milky Way halo along a large number of sightlines (Richter et 
al.\,2005; Ben Bekhti et al.\,2008,2011). As an example 
for these data, we show in Fig.\,1 the velocity profile of 
Ca\,{\sc ii} $\lambda 3934.8$ absorption in the VLT/UVES 
spectrum of the QSO PKS\,1448$-$232. Apart from the absorption 
related to the local disk of the Milky Way in the range 
$v_{\rm LSR}=-30$ to $+50$ km\,s$^{-1}$, high-velocity 
Ca\,{\sc ii} absorption is visible between $v_{\rm LSR}=-90$ 
and $-180$ km\,s$^{-1}$, originating in neutral gas 
in an HVC in the Milky Way halo, located several kpc above 
the Galactic plane (see Richter et al.\,2005). In Ben Bekhti 
et al.\,(2008,2011) we have analyzed several hundred of 
such Ca\,{\sc ii} spectra and have studied the velocity 
distribution of extra-planar neutral gas structures in the 
Milky Way halo. We find that the majority of the Ca\,{\sc ii} 
halo absorbers have radial velocities in the range 
$|v_{\rm LSR}|\leq 200$ km\,s$^{-1}$, but there is an 
excess of absorbers with negative radial velocities. 
A similar trend is observed in the UV using Si\,{\sc ii} as
tracer for neutral gas in the Milky Way halo (Herenz et al.\,2013).
This excess possibly reflects a net-infall of neutral gas towards
the Milky Way as part of the on-going gas accretion of the
Galaxy (Richter 2012). 

% ---------------- Fig 2
\begin{figure}
\includegraphics[angle=0, width=\linewidth]{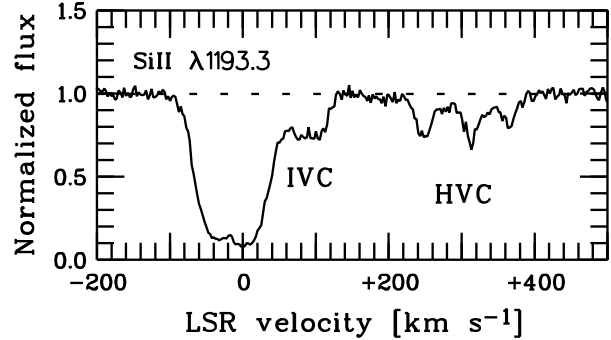}
\caption{The velocity profile of Si\,{\sc ii} $\lambda 1193.3$ 
absorption in the \emph{HST}/COS spectrum of the Seyfert 1 
galaxy IRAS\,F04250$-$5718 indicates a complex absorption
pattern with various IVC and HVC velocity components.} 
\label{Flabel2}
\end{figure}

Based on UV observations of five-times ionized oxygen (O\,{\sc vi}) with the
\emph{Far Ultraviolet Spectroscopic Explorer} (\emph{FUSE}),
Sembach et al.\,(2003) have shown that the velocity distribution 
of the highly-ionized circumgalactic gas component of the Milky Way 
extends to substantially higher radial velocities, with many 
O\,{\sc vi} absorbers located in the velocity range 
$|v_{\rm LSR}|=200-400$ km\,s$^{-1}$. The (on average) higher
radial velocities of circumgalactic O\,{\sc vi} compared to 
Ca\,{\sc ii} and Si\,{\sc ii} most likely indicate that the 
highly-ionized gas phase in the CGM is spatially more extended 
than the neutral phase. For Milky Way-type galaxies, that do not 
drive massive winds of ionized gas into their circumgalactic 
environment, such a trend is not surprising: as the gas approaches 
the central region of the Galaxy's potential well, the gas pressure 
increases, while the ionization fraction is expected to decrease
because of enhanced cooling and recombination. Constrained 
simulations of the Milky Way and its gaseous environment 
further support this scenario (Nuza et al.\,2013; Marasco
et al.\,2013).

The velocity pattern of Galactic IVCs and HVCs can vary
dramatically, reflecting the inhomogeneous space distribution
and radial-velocitiy distribution of gas structures and satellite
galaxies in the circumgalactic environment of the Milky Way and
in the Local Group. This aspect is demonstrated in Fig.\,2,
where we show the complex velocity profile of Si\,{\sc ii} absorption
in the direction of the Seyfert 1 galaxy IRAS\,F04250$-$5718,
based on data obtained with the Cosmic Origins Spectrograph
(COS) onboard the \emph{Hubble Space Telescope} (\emph{HST}).
At galactic coordinates $l=267$ deg and $b=-42$ deg, this sightlines passes
an IVC in the inner Milky Way halo and the so-called Magellanic Stream
(at $|v_{\rm LSR}|\geq 200$ km\,s$^{-1}$; see Sect.\,2.3). In
addition, some of the high-velocity Si\,{\sc ii} absorption
possibly originates in the extended ionized gaseous envelope of
the Large Magellanic Cloud (LMC).

\subsection{Characteristics of optical and UV spectra}

Most of the low, intermediate and high metal ions have their
transitions in the UV range. With sufficiently bright UV
background sources (such as QSOs), the absorption lines
arising from the transitions can be used very efficiently 
to explore the physical nature and chemical composition of 
gaseous material that is located between us and the background 
source. Among the most powerful diagnostic lines in the UV, 
that are commonly used to study cold, warm, and hot gas
in the ISM, CGM, and IGM, are the lines from 
C\,{\sc ii} $\lambda 1334.5$,
C\,{\sc iv} $\lambda\lambda 1548.2,1550.8$,
O\,{\sc i} $\lambda 1302.2$,
O\,{\sc vi} $\lambda\lambda 1031.9,1037.6$,
Si\,{\sc ii} $\lambda\lambda 1190.4,1193.3,1260.4,1304.4,1526.7$,
Si\,{\sc iii} $\lambda 1206.5$,
Si\,{\sc iv} $\lambda\lambda 1393.8,1402.8$,
Mg\,{\sc ii} $\lambda\lambda 2796.4,2803.5$, and
Fe\,{\sc ii} $\lambda\lambda 1144.9,1608.5,2586.7,2600.2$.
Despite the importance of the UV range for our understanding 
of the diffuse gas component in the local Universe,
current UV spectra of extragalactic (and Galactic) background
sources are limited in spectral resolution and signal-to-noise 
(S/N) simply due to the fact that space-based observatories are 
required to obtain these data. With a resolving power of up to 
$R=114,000$, the Space Telescope Imaging Spectrograph (STIS) 
onboard \emph{HST} currently is
the UV spectrograph with the highest spectral resolution, but
only a very limited number of extragalactic background sources
can be observed with STIS at a S/N of $>30$ and 
a typical spectral resolution of $R=45,000$. Another powerful 
UV spectrograph onboard \emph{HST} is COS, which 
covers a NUV and FUV wavelength range similar to that
of STIS,  but which has a substantially higher sensitivity. 
Thus, COS is able to observe hundreds of extragalactic targets 
at high S/N, yet the spectral resolution
is limited to $R\leq 24,000$.

To overcome the problem of lack in spectral resolution and to minimize
the resulting systematic errors for the determination of
column densities and absorber kinematics, supplementary 
optical spectra can be used. The optical regime covers only a 
very limited number of useful transitions of metal ions. These 
transitions are predominantly from neutral and singly-ionized 
species (e.g., Na\,{\sc i} $\lambda\lambda 5891.6,5897.6$
Ca\,{\sc ii} $\lambda\lambda 3934.8,3969.6$,
Ti\,{\sc ii} $\lambda 3384.7$; but see Sect.\,2.4).
Optical spectrographs installed on $8-10$m-class
telescopes on the ground are, however, able to deliver 
hundreds of high-S/N spectra (S/N\,$>30$) even at high 
spectral resolution ($R\geq 40,000$; see Fig.\,1). 
Such spectra provide a deep insight into the velocity-component 
structure and gas kinematics of the absorbers 
(e.g., Welty et al.\,1999) and thus can complement 
lower-resolution UV spectra that deliver detailed 
information on the chemical composition and the physical 
conditions of the gas.

In the following, we provide an example for the combined
use of optical and UV spectral data to explore the nature
of the Milky Way's circumgalactic medium.

\subsection{A prime example: absorption spectroscopy of 
the Magellanic Stream}

One of the most prominent circumgalactic gas features of the 
Milky Way is the Magellanic Stream (MS). The Stream represents a massive
($\sim 10^8-10^9\,M_{\sun}$) gaseous structure consisting of both neutral
and ionized gas and is believed to be a result of the interaction of the two
Magellanic Clouds as they approach the Milky Way (e.g., Wannier \& Wrixon 1972; 
Gardiner \& Noguchi 1996; Weiner \& Williams 1996; Putman et al.\,2003; 
Br\"uns et al.\,2005). Although the MS has
a relatively large distance of $\sim 50-60$ kpc, it covers about
4 percent of the 21cm high-velocity sky (see Wakker 2004).
The Stream is known to contain heavy elements as well as dust grains and 
molecular hydrogen (Gibson et al.\,2000; Sembach et al.\,2001; 
Richter et al.\,2001a; Fox et al.\,2013; Richter et al.\,2013).
Ever since its detection more than 40 years ago it was speculated
that the MS has its origin in one {\it or} the other Magellanic Cloud.

Because of the relatively large angular extent of the MS, there are a number 
of bright background quasars behind the Stream that can be used to study
the chemical composition of the MS by way of absorption spectroscopy
(e.g., Fox et al.\,2005,2010). Tidal models indicate that the
main body of the Stream most likely was stripped from its parent galaxy 
$\sim 1-2$ Gyr ago (e.g., Gardiner \& Noguchi 1996; Connors et al.\,2006; 
Besla et al.\,2010,2012) and the MS does not contain any 
stars that could have influenced its metallicity since then. Therefore, 
the present-day abundance of heavy elements in the MS represents the 
chemical composition of its parent galaxy $\sim 1-2$ Gyr ago. From the 
analysis of a UV spectrum of the Seyfert\,1 galaxy Fairall\,9 obtained with 
the Goddard High Resolution Spectrograph (GHRS) onboard the 
HST, together with Parkes 21\,cm H\,{\sc i} data, 
Gibson et al.\,(2000) obtained a metallicity of the Stream 
of [M/H$]=-0.55\pm0.06^{+0.17}_{-0.21}$ ($\sim 0.3$ solar). Although
this measurement represented a major step in our understanding of
the Stream's chemical composition, it was not accurate enough
to clearly place the origin of the MS in either the SMC or the LMC.

% ---------------- Fig 3
\begin{figure}
\includegraphics[angle=0, width=\linewidth]{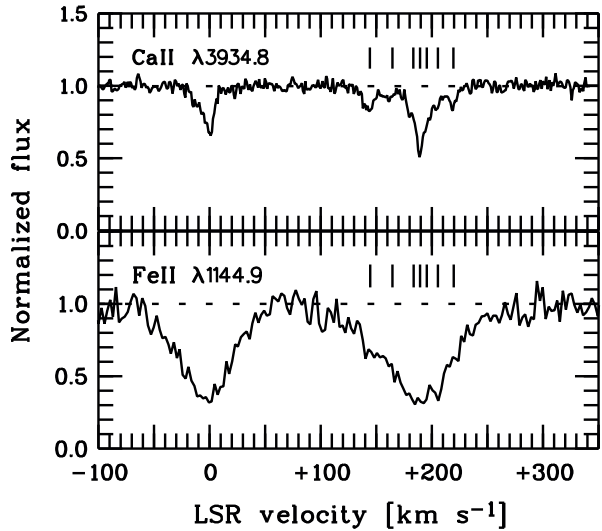}
\caption{
Velocity-component structure of the gas in the MS in 
the Ca\,{\sc ii} $\lambda 3934.8$ line (upper panel), based on
the high-resolution ($R=70,000$) UVES spectrum of Fairall\,9.
Fe\,{\sc ii} $\lambda 1144.9$ absorption in the lower-resolution 
COS spectrum ($R=16,000$) of Fairall\,9 is
shown for comparison (lower panel).}
\label{Flabel3}
\end{figure}

To further increase the accuracy of the metallicity determination of the
Stream towards Fairall\,9 and other background targets and to study in
detail the chemical composition of the gas and dust in the MS, 
more accurate spectral data are needed. As part of our ongoing project 
to study the properties of the Magellanic Stream in absorption, we 
recently have obtained for six lines of sight (including the 
Fairall\,9 sightline) high-resolution optical data from VLT/UVES
and medium-resolution UV data from \emph{HST}/COS, all data sets providing
absorption spectra with excellent S/N ratios (Fox et al.\,2013). 
Only the combination of optical data with high
spectral resolution and UV data with high S/N ratio enables us to fully
resolve the Stream's velocity-component structure and to substantially
minimize the systematic errors in the determination of metal column densities,
because the MS is a massive gas cloud with complex internal kinematics 
(e.g., Nidever et al.\,2008). 

For the Fairall\,9 sightline, the MS Ca\,{\sc ii} absorption pattern 
seen in the optical UVES data indicates a complex internal velocity 
structure of the Stream in this direction with seven individual
absorption components within a velocity range of $\Delta v=80$ km\,s$^{-1}$
centered at $v_{\rm LSR}\sim 190$ km\,s$^{-1}$ (Richter et al.\,2013). 
In Fig.\,3, upper panel, we show the velocity-component structure of 
the gas in the MS in the Ca\,{\sc ii} $\lambda 3934.8$ line based on 
the high-resolution ($R=70,000$) UVES spectrum of Fairall\,9. 
For comparison, Fe\,{\sc ii} $\lambda 1144.9$ absorption 
in the lower-resolution COS spectrum ($R=16,000$) of Fairall\,9 is 
shown in the lower panel of Fig.\,3.

With this new spectral information at hand, we determine a surprisingly
high metal abundance in the Stream along the Fairall\,9 sightline. 
Using the unsaturated S\,{\sc ii} absorption in the \emph{HST}/COS data, 
in combination with H\,{\sc i} 21cm data from the Galactic-All Sky Survey 
(GASS; McClure-Griffiths et al.\,2009, Kalberla et al.\,2010),
the $\alpha$ abundance in the MS turns out to be [S/H$]=-0.30\pm0.04$ 
($0.50$ solar). This value is $\sim 70$ percent higher than
the previous estimate for [S/H] in the MS towards Fairall\,9
presented by Gibson et al.\,(2000; see Richter et al.\,2013 for 
a discussion on the origin for this discrepancy).
Most striking, this $\alpha$ abundance is five-times higher than what is 
found along the other five MS sightlines presented in Fox et al.\,(2013)
based on COS/UVES data sets with similar data quality.
The substantial differences in the chemical composition of the MS
towards Fairall\,9 compared to the other sightlines suggest that
the enrichment history of the Stream is far more complex than
previously thought. Most likely, the main body of the MS, that
has an $\alpha$ abundance of $\sim 0.1$ solar (Fox et al.\,2013),
originates in the SMC, from which the bulk of the Stream's material
was separated $\sim 1-2$ Gyr ago. In contrast, the metal-rich
gas towards Fairall\,9 stems from the LMC, where the
gas was locally enriched by a massive star burst, blown away from the 
galaxy, and then incorporated into the Stream. 

The existence of a metal-enriched filament in the Stream towards
Fairall\,9 that originates in the LMC is supported by a kinematic study
of the MS from Nidever et al.\,(2008), who performed a 
systematic Gaussian decomposition of the Stream's H\,{\sc i} velocity 
profiles using the LAB 21cm all-sky survey (Kalberla et al.\,2005). 

\subsection{[FeX] absorption as possible tracer for a warm-hot CGM}

The resonance lines of atomic and molecular species located in the UV 
serve as tracer of cold, warm, and hot gas in a temperature range of
$T\approx30-300,000$ K. For collisionally ionized gas, 
O\,{\sc vi}, with an ionization potential of creation of 
$E=114$ eV and its two strong transitions at $1031.9$ and
$1037.6$ \AA, is the best ion in the UV to study the warm-hot 
CGM around the Milky Way at $T\approx 300,000$ K. UV observations 
carried out with {\it FUSE} have demonstrated that high-velocity O\,{\sc vi} 
absorption is present in many directions of the sky with a large 
covering fractions of $f_{\rm c}\approx 0.6$ for column densities
log $N$(O\,{\sc vi}$)\geq 13.4$ (Sembach et al.\,2003). The O\,{\sc vi}
absorption possibly indicates transition-temperature gas in the 
interfaces between cooler gas clouds and a surrounding hot coronal 
gas (see also Fox 2011).

Numerical simulations of large-scale structure formation 
predict that the dominant fraction of the baryonic matter in the 
IGM and CGM resides in a hot phase at million-degree temperatures 
(e.g., Cen \& Ostriker 1999; Dav\'e et al.\, 2001; 
Tepper-Garc\'ia et al.\,2011).
The high temperature of this gas phase (commonly referred to 
as warm-hot intergalactic medium, WHIM) arises from shock-heating 
of collapsing large-scale strucures. Because the cooling-time of 
the WHIM is on the order of the Hubble-time, simulations indicate 
that there exists a widespread, highly-ionized CGM gas phase 
around galaxies that cannot be readily observed in the UV due to 
the lack of appropriate transitions of high ions that have 
ionization potentials beyond that of O\,{\sc vi}.

X-ray spectra of bright Blazars, obtained with Chandra and XMM-Newton, 
have been used to search for this high-temperature gas phase through 
line absorption in the higher ionization states of oxygen, O\,{\sc vii} 
and O\,{\sc viii} (e.g., Gupta et al.\,2012). Following these authors, 
the total mass in the warm-hot CGM around the Milky Way may be as large 
as $\sim 6\times 10^{10}$ M$_{\sun}$ (but see Bregman \& Lloyd-Davies
2007). Due to the severe limitations in 
spectral resolution and S/N, however, the interpretation of the X-ray 
O\,{\sc vii}/O\,{\sc viii} absorption features with respect to origin, 
mass, and physical nature of the gas is afflicted with large systematic 
uncertainties (see Richter et al.\,2008 for a more detailed discussion 
on this topic). 

For a better understanding of the properties of warm-hot gas 
in the outer environment of galaxies it is of prime importance to search 
for alternative ways to trace hot gas in the Universe through absorption 
and to combine results from independent observational methods and from 
simulations. In search for strategies to detect million-degree gas 
throughout the Universe, York \& Cowie (1983) discussed the possibility
to use the optical intersystem lines of [Fe\,{\sc x}] $\lambda 6374.5$ and 
[Fe\,{\sc xiv}] $\lambda 5302.9$ as spectral tracers for million-degree gas. 
These lines have oscillator strengths as low as $f_{\lambda 6374}=2.1 \times
10^{-7}$ and $f_{\lambda 5302}=5.1 \times 10^{-7}$ and thus an
extremely high S/N in optical spectra is required to detect
these lines in the presence of hot, collisionally ionized gas.
Note that it requires energies of $234$ eV to ionize Fe from 
Fe$^{+8}$ to Fe$^{+9}$ and $361$ eV from Fe$^{+13}$ to Fe$^{+14}$ 
(e.g., Sutherland \& Dopita 1996 and references therein).
Because the UV background at $z=0$ does not provide sufficient
photons at such high energies, collisional ionization is the only 
mechanism that is expected contribute to a possible population of
Fe$^{+9}$ and Fe$^{+13}$ ions in intergalactic and 
circumgalactic environments.

In a series of papers, Hobbs (1984a, 1984b, 1985), Hobbs \& Albert (1985),
and Pettini \& D'Odorico (1986) searched for [Fe\,{\sc x}] and [Fe\,{\sc xiv}] 
absorption in the spectra of stars, but without success. Using the optical 
spectrum of the extremely bright supernova SN\,1987A in the LMC, several
groups (Wang et al.\,1989; Pettini et al.\,1989; Malaney \& Clampin 1989)
detected a weak absorption ($W_{\lambda}\sim 3$ m\AA) feature near $6380$ \AA\, 
that they identified as Doppler-shifted [Fe\,{\sc x}] $\lambda 6374.5$ absorption 
originating in hot gas in the direction of the LMC. This feature, if correctly 
identified, would imply a very large column of ionized gas in this direction of 
log $N$(H\,{\sc ii})$\approx21.5$ and would indicate the existence of a huge 
reservoir of baryonic matter hidden in hot gas. The interpretation of the 
observed absorption feature near $6380$ \AA\, in the spectrum of SN\,1987A as [Fe\,{\sc x}] 
absorption was disputed by Wampler et al.\,(1991), however, who attributed the
observed feature to a Doppler-shifted diffuse interstellar band (DIB). Indeed, 
various DIBs and telluric lines strongly hamper a definite identification 
of [Fe\,{\sc x}] absorption in this wavelength region (see, e.g., Jenniskens 
\& D\'esert 1994). Since these studies of the optical spectrum of SN\,1987A 
no further attempts (that we are aware of) have been made to search for hot 
gas in the CGM and IGM using the [Fe\,{\sc x}] $\lambda 6374.5$ transition.

% ---------------- Fig 4
\begin{figure}
\includegraphics[angle=0, width=\linewidth]{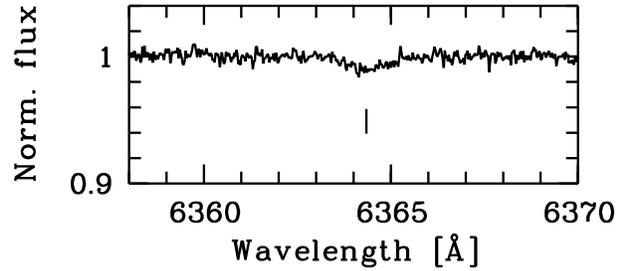}
\caption{
The VLT/UVES spectrum of the bright quasar HE\,0515$-$4414 in the
wavelength range between $6358$ and $6370$ \AA\, is shown.
A very weak, broad absorption feature is seen near $6364$ \AA\,
corresponding to a LSR velocity of $-470$ km\,s$^{-1}$ if
the feature would be Doppler-shifted [Fe\,{\sc x}] $\lambda 6374.5$
absorption. The origin of this feature remains unclear, however.}
\label{Flabel4}
\end{figure}

In an on-going long-term project that aims at studying optical absorption 
features in circumgalactic gas of the Milky Way and other galaxies using 
archival VLT/UVES data (Richter et al.\,2005; Ben Bekhti et al.\,2008,2011;
Richter et al.\,2011), we have revisited the issue of 
[Fe\,{\sc x}] and [Fe\,{\sc xiv}] absorption from a possibly existing 
hot gas component in the CGM. While this project is on-going, we here 
present some preliminary results and shortly outline our analysis strategy.

The majority of the more than 500 high-resolution ($R>40,000$) VLT/UVES 
QSO spectra that are available to us in the archive have a S/N of $<150$ per 
resolution element. Only a few spectra in our VLT/UVES sample have a S/N 
near $6375$ \AA\, that is substantially higher than that, so that the 
direct search for [Fe\,{\sc x}] $\lambda 6374.5$ and [Fe\,{\sc xiv}] 
$\lambda 5302.9$ absorption along individual sightlines is limited to a 
few cases. To visualize the expected absorption characteristics
of [Fe\,{\sc x}] absorption we show in Fig.\,4 the VLT/UVES spectrum of 
the bright quasar HE\,0515$-$4414 in the wavelength range between $6358$ and $6370$
\AA. In this spectrum, the S/N is as high as $530$ per resolution element 
at $6375$ \AA. A very weak, broad absorption feature is visible near 
$6364$ \AA, corresponding to a LSR velocity of $-470$ km\,s$^{-1}$ if 
the feature were Doppler-shifted [Fe\,{\sc x}] $\lambda 6374.5$ absorption. 
On a first glance, this absorption feature appears to be a promising
candidate for [Fe\,{\sc x}] absorption from hot gas in the 
Local Group. With a FWHM of 
$\sim 65$ km\,s$^{-1}$ it's shape matches the expected large width of a 
Fe absorption line arising in million-degree circumgalactic gas, 
in which both thermal and non-thermal line broadening mechanisms are 
relevant. Moreover, the observed absorption feature is most likely not 
related to a DIB or a telluric line. However, 
with an equivalent width of almost $13$ m\AA\, the absorption is 
{\it too strong} to realistically represent [Fe\,{\sc x}] absorption 
in local warm-hot circumgalactic gas, as it would trace an ionized gas column
of log $N$(H\,{\sc ii}$)>22.2$ for collisionally ionized gas at a temperature of 
$T\approx 10^6$ K. Numerical simulations of Milky Way-type galaxies 
and their gaseous environment (e.g., Nuza et al.\,2013) suggest instead that the 
column densities of hot, ionized gas in the CGM are below 
log $N$(H\,{\sc ii}$)=21$. [Fe\,{\sc x}] $\lambda 6374.5$ absorption from 
warm-hot gas in the Milky Way environment thus is expected to be subtantially 
weaker than the feature observed towards HE\,0515$-$4414. 

In order to reach the very high S/N of $>1000$ that is necessary to 
detect such extremely weak [Fe\,{\sc x}] lines the stacking of a large 
number of QSO spectra is required. We are currently 
working on such a stacking procedure using those VLT/UVES data sets that cover 
the [Fe\,{\sc x}] wavelength range at a S/N ratio that allows us to safely 
remove DIBs, telluric lines, and other {\it known} blending features 
before the data is used in the stacking routine. In addition to the 
high-resolution UVES data, we are also planning to stack thousands of 
low-resolution spectra from the Sloan Digital Sky Survey (SDSS) to 
constrain the amount of hot gas in the Local Group using the
[Fe\,{\sc x}] $\lambda 6374.5$ transition. The first 
results from this project will be presented in a forthcoming paper
(Richter et al.\,2014).

\section{Concluding remarks}

High-resolution absorption spectroscopy is the most 
efficient method to explore the nature of diffuse
intergalactic and circumgalactic gas in the local Universe.
As we have adumbrated in this article,
spectroscopic observations of the circumgalactic 
medium of the Milky Way indicate a high complexity 
of the gas kinematics and a large range in physical 
conditions in such gas. 
Future spectroscopic observations of the CGM at low
$z$ that combine the strengths of optical and UV
instruments and that also take into account the results
from other wavelength regimes (e.g., infrared, radio, X-ray)
will be of great importance to further constrain the role
of the circumgalactic medium for the ongoing formation and evolution
of galaxies and to characterize its connection to the cosmic web.

\acknowledgements

Support for this research was provided by NASA through
grant HST-GO-12604 from the Space Telescope Science Institute,
which is operated by the Association of Universities for Research
in Astronomy, Incorporated, under NASA contract NAS5-26555.

% Use this code if you wish to generate your bibliography with BibTeX;
% please replace first the string "an-demo" below with the name(s) of
% the BibTeX data base(s) you want to use.
% The resulting bibliography-output (the contents of the .bbl file)
% must be pasted into this file before submission.
%
% \bibliographystyle{an}
% \bibliography{an-demo}
%
% Replace the following example bibliography with your references
% before submission:

\end{document}